# Evaluating IEEE 802.15.4 for Cyber-Physical Systems

Feng Xia[1], Alexey V. Vinel[2], Ruixia Gao[1], Linqiang Wang[1], Tie Qiu[1]
[1]School of Software, Dalian University of Technology, Dalian 116620, China
e-mail: f.xia@ieee.org
[2]Tampere University of Technology, Finland / SPIIRAS, Russia
e-mail: vinel@ieee.org

**Abstract**

*With rapid advancements in sensing, networking, and computing technologies, recent years have witnessed the emergence of cyber-physical systems (CPS) in a broad range of application domains. CPS is a new class of engineered systems that features the integration of computation, communications, and control. In contrast to general-purpose computing systems, many cyber-physical applications are safety-cricial. These applications impose considerable requirements on quality of service (QoS) of the employed networking infrastruture. Since IEEE 802.15.4 has been widely considered as a suitable protocol for CPS over wireless sensor and actuator networks, it is of vital importance to evaluate its performance extensively. Serving for this purpose, this paper will analyze the performance of IEEE 802.15.4 standard operating in different modes respectively. Extensive simulations have been conducted to examine how network QoS will be impacted by some critical parameters. The results are presented and analyzed, which provide some useful insights for network parameter configuration and optimization for CPS design.*

## 1. Introduction

There is a revolutionary transformation from stand-alone embedded systems to networked cyber-physical systems (CPS) that bridge the virtual world of computing and communications and the real world [1-3]. Cyber-physical systems are tight integrations of computation, networking, and physical objects, in which embedded devices are networked to sense, monitor and control the physical world. CPS is rapidly penetrating every aspect of our lives and plays an increasingly important role. This new class of engineered systems promises to transform the way we interact with the physical world just as the Internet transformed how we interact with one another. Before this vision becomes a reality, however, a large number of challenges have to be addressed, including e.g. resource constraints, platform heterogeneity, dynamic network topology, and mixed traffic [4]. High-confidence wireless communication protocol design in the context of CPS is among those issues that deserve extensive research efforts.

The IEEE 802.15.4 protocol [5] is a low-rate, low-cost, and low-power communication protocol for wireless interconnection of fixed and/or portable devices. Currently it has become one of the most popular communication standards used in the field of wireless sensor networks (WSNs). On the other hand, cyber-physical systems are generally built upon wireless sensor and actuator networks (WSANs), which is an extension of WSNs. In this context, WSANs are generally responsible for information exchange (i.e. data transfer), serving as a bridge between the cyber and the physical worlds. As a consequence, the IEEE 802.15.4 protocol will be utilized in many cyber-physical systems and applications of today and tomorrow. Despite the wide popularity of IEEE 802.15.4 networks, their applicability to CPS needs to be validated. This is because IEEE 802.15.4 was not designed for networks that can provide quality of service (QoS) guarantees, while the performance of cyber-physical applications often depend highly on the QoS of underlying networks. Therefore, it becomes necessary and important to evaluate the performance of IEEE 802.15.4 protocol in the context of CPS, which forms the focus of this paper.

IEEE 802.15.4 supports two basic kinds of networking topologies relevant to CPS applications: star and peer-to-peer. Since most CPS applications involve monitoring tasks and reporting towards a central sink, here we focus on a one-hop star network. All the nodes are set to be in each other's radio range. Consequently, there are no hidden nodes. IEEE 802.15.4 medium access control (MAC) adopts carrier sense multiple access with collision avoidance (CSMA/CA) as the channel access mechanism. In an IEEE 802.15.4-based one-hop star network, the network QoS in terms of e.g. packet loss rate and latency depends on the number of nodes competing for channel access and their packet generation rates as well as the configuration of MAC parameters in the nodes. The IEEE 802.15.4 specification suggests default values for different MAC parameters. However, as demonstrated later in this paper, the default configuration may not necessarily yield the best QoS in all situations with different traffic load. In fact, it is very difficult, if not impossible, to determine a single IEEE 802.15.4 MAC configuration that always results in the optimal performance, which will be supported by our results.

In this paper, we will evaluate the performance of IEEE 802.15.4 protocol in both beacon-enabled and non-beacon-enabled modes respectively. We consider a star network of several nodes collecting data and transmitting them to a central sink node. The network QoS is characterized by several metrics, including effective data rate, packet loss rate, and average end-to-end delay. These metrics will be examined with respect to different MAC parameter settings. We contribute to better understanding of the IEEE 802.15.4 standard in the context of CPS by presenting a set of results of simulation experiments using OMNeT++, which is a popular open-source simulation platform especially suitable for simulation of communication networks.

The remainder of this article is organized as follows. Section 2 gives an overview of related work in the literature. In Section 3 we discuss the major features of CPS and their requirements on QoS from a networking perspective. In Section 4, we introduce briefly the IEEE 802.15.4 standard. This is followed by a description of simulation settings including simulation scenario and parameter settings, and a definition of performance metrics in Section 5. Section 6 and 7 present and analyze the simulation results. Finally, Section 8 concludes the paper.

## 2. Related Work

CPS has been attracting rapidly growing attention from academia, industry, and the government worldwide. A number of conferences, workshops, and summit on CPS have been held during the past several years, gathering researchers, practitioners, and governors from all around the world to discuss the challenges and opportunities brought by CPS. The renowned CPS Week launched in 2008 and is held annually. Many world-leading IT companies such as Microsoft, IBM, National Instruments, NEC Labs, and Honeywell have started research and development initiatives closely related to CPS. Although there have been a lot of research results in related fields including embedded computing systems, ubiquitous computing, and wireless sensor networks, CPS is a relatively new area with a large number of open problems [1]. In particular, we pay special attention to performance evaluation of one of the most popular wireless communication protocols (i.e. IEEE 802.15.4) in the context of CPS.

Since the release of IEEE 802.15.4 in 2003 and the emergence of the first products on the market there have been many analytical and simulation studies in the literature, trying to characterize the performance of the IEEE 802.15.4 standard [6,7]. However, most of these studies mainly focus on IEEE 802.15.4 in either the beacon-enabled mode or the non-beacon-enabled mode. For example, Lu et al [8] conducted performance evaluation of IEEE 802.15.4 using the NS-2 network simulator, focusing on its beacon-enabled mode for a star-topology network. Pollin et al [9] provided an analytical Markov model that predicts the performance and detailed behavior of the IEEE 802.15.4 slotted CSMA/CA mechanism. Jung et al [10] enhanced Markov chain models of slotted CSMA/CA IEEE 802.15.4 MAC to account for unsaturated traffic conditions. Huang et al [11] and Ren et al [12] focused on analyzing beacon-enabled IEEE 802.15.4 network by setting two system parameters, i.e. Beacon Order and Superframe Order. In [13] Buratti established a flexible mathematical model for beacon-enabled IEEE 802.15.4 MAC protocol in order to study beacon-enabled 802.15.4 networks organized in different topologies.

On the other hand, many works on IEEE 802.15.4 are based on non-beacon-enabled mode [14,15]. In [16], for example, Latre et al studied the performance of the non-beaconed IEEE 802.15.4 standard in a scenario containing one sender and one receiver. In [17], Rohm et al analyzed via simulations the impact of different configurable MAC parameters on the performance of beaconless IEEE 802.15.4 networks under different traffic loads. In [18], Rohm et al measured the performance of beaconless IEEE 802.15.4 networks with various system parameters under different traffic load conditions. Buratti and Verdone [19] provided an analytical model for non-beacon-enabled IEEE 802.15.4 MAC protocol in WSN, which allows evaluation of the statistical distribution of traffic generated by nodes.

In addition, many researchers have studied IEEE 802.15.4 for special application environments [20,21]. In [22], Chen et al analyzed the performance of beacon-enabled IEEE 802.15.4 for industrial applications in a star network in OMNeT++. The effects of varying the payload size, sampling and transmitting cycles in an IEEE 802.15.4 based star network that consists of ECG monitoring sensors are analyzed in [23]. Li et al [24] studied the applicability of IEEE 802.15.4 over a wireless body area network by evaluating its performance. In [25], Liu et al paid attention to study the feasibility of adapting IEEE 802.15.4 protocol for aerospace wireless sensor networks. By analyzing the IEEE 802.15.4 standard in a simulation environment, Chen et al [26] modified IEEE 802.15.4 protocol for real-time applications in industrial automation. Mehta et al [27] proposed an analytical model to understand and characterize the performance of GTS traffic in IEEE 802.15.4 networks for emergency response. In [28], Zen et al analyzed the performance of IEEE 802.15.4 to evaluate the suitability of the protocol in mobile sensor networking.

In this paper, we extend our previous work [29]. There are two key contributions. First, we comprehensively study the performance of IEEE 802.15.4 protocol in both beacon-enabled and non-beacon-enabled modes based on a one-hop star network, using the OMNeT++ simulator. We select end-to-end delay, effective data rate and packet loss rate as the network QoS metrics and analyze how they will be affected by several important protocol parameters. Second, we make an in-depth analysis of the results to provide insights for adapting IEEE 802.15.4 for CPS. By analyzing the results, we can configure and optimize the parameters of IEEE 802.15.4 for CPS.

## 3. QoS Requirements of CPS

As mentioned previously, CPS is a new class of systems of systems that tightly integrate computation, networking, and physical objects. They feature by nature the convergence of computing, communications, and control (i.e. 3C). In a feedback manner, the cyber world and the physical world exchange information and effect on each other, thus forming a closed-loop system. The basic goal of CPS is to sense, monitor, and control physical environments/objects effectively and efficiently.

A typical CPS mainly consists of the following components: physical objects, sensors, actuators, communication networks, and computing devices (e.g. controllers). Various sensors and actuators will be geographically distributed and directly coupled with physical objects. Sensors collect the state information of physical objects and send it to certain computing nodes through the communication networks. The network could possibly be a combination of multiple networks of different types, e.g. wired and wireless networks. It is responsible for transferring data reliably and in real-time. Relatively complex decision-making algorithms will be generally executed on computing devices, which generate control commands based on information collected by sensors. In practice, these computations could be completed in a distributed or centralized manner. The control commands will then be sent to actuators, also via the networks if needed, and be performed by the appropriate actuators. In this way, CPS facilitates interplay of the cyber and physical systems, i.e. control of physical environments.

As we can see, cyber-physical systems in general are built on WSANs, though the networks within a real-world CPS could potentially be much more complex and heterogeneous. Particularly, when the scale of a CPS becomes very large, WSAN is a natural choice for interconnection of a large number of sensor, computing, and actuator nodes due to the celebrated benefits of wireless networking (as compared to wired counterparts). The use of WSAN distinguishes CPS from traditional embedded systems and wireless sensor networks. From a networking viewpoint, some widely-recognized characteristics of CPS can be outlined as follows.

1) *Network complexity*. Due to various reasons, such as different node distances, diverse node platforms and operating conditions, multiple communication networks of different types could be employed in one single CPS. Different communication protocols/standards may co-exist. The network of a typical CPS is often large in scale because of the large number of distributed nodes in the systems.
2) *Resource constraints*. In CPS, cyber capabilities are embedded into physical objects/nodes. These embedded devices are always limited in computing speed, energy, memory, and network bandwidth, etc. For example, for an IEEE 802.15.4 network, the bandwidth is limited to 250 kbps.
3) *Hybrid traffic and massive data*. In a large-scale CPS, diverse applications may need to share the same network, causing mixed traffic. The large number of sensor and computing nodes generate a huge volume of data of various types. In particular, in order to sense the state of physical world correctly and accurately, a CPS usually needs to collect a mass of data by using diverse sensors. This data must be processed and transmitted properly.
4) *Uncertainty*. In CPS there are many factors that could potentially cause uncertainty with various attributes, including e.g. sensor measurement error, computational model error, software defect, environmental noise, unreliability of wireless communications, and changes in network topology (due to e.g. node failure or mobility).

CPS can be applied in a wide range of domains. Potential applications of CPS include assisted living, integrated medical systems, safe and efficient transportation, automated traffic control, advanced automotive systems, autonomous search and rescue, energy conservation, energy efficient buildings, environmental control, factory automation, home automation, critical infrastructure control, distributed autonomous robotics, defense, etc. Ubiquitous applications and services that could significantly improve the quality of our daily lives will be enabled by CPS, which will make applications more effective and more efficient. However, the success of these applications heavily relies on the QoS provided by the employed networks. Therefore, WSANs for CPS have to deliver massive data within hybrid traffic in a proper manner with the presence of network complexity, resource constraints, and uncertainty. Particularly, in most CPS applications the network QoS needs to satisfy the requirements on several non-functional properties, i.e., real-time, reliability, and resource efficiency [4,30,31]. Based on this observation, in this paper we focus our attention to examine the capability of IEEE 802.15.4 in guaranteeing QoS in terms of these properties.

## 4. IEEE 802.15.4 Standard

In this section we give a brief introduction to the IEEE 802.15.4 protocol specification for the sake of integrality. More details of the standard can be found in [5]. The specification defines the physical (PHY) and MAC layer. The PHY layer is defined for operation in three different unlicensed ISM frequency bands (i.e. the 2.4 GHz band, the 915 MHz band, and the 868 MHz band) that include totally

27 communication channels. An overview of their modulation parameters is shown in Table 1.

Table 1. **IEEE 802.15.4 frequency bands**

| Frequency (MHz) | Frequency band (MHz) | Data rate (kbps) | Modulation scheme | Operating region |
|---|---|---|---|---|
| 868 | 868-868.6 | 20 | BPSK | Europe |
| 915 | 902-928 | 40 | BPSK | North America |
| 2400 | 2400-2483.5 | 250 | O-QPSK | Worldwide |

There are two different kinds of devices defined in IEEE 802.15.4: full function device (FFD) and reduced function device (RFD). An FFD can act as an ordinary device or a PAN coordinator. But RFD can only server as a device supporting simple operations. An FFD can communicate with both RFDs and other FFDs while an RFD can only communicate with FFDs.

IEEE 802.15.4 supports a star topology or a peer-to-peer topology. In star networks, all the communications are between end devices and the sink node which is also called PAN coordinator. The PAN coordinator manages the whole network, including distributing addresses to the devices and managing new devices that join in. In the peer-to-peer network, the devices can communicate with any other devices which are within their signal radiation ranges. A specific type of peer-to-peer networks is cluster tree networks. In this case, most of the devices are FFD. RFD can only communicate with one FFD sometime.

### 4.1 Superframe Structure

The IEEE 802.15.4 standard allows two kinds of network configuration modes:
1) *Beacon-enabled mode:* a PAN coordinator periodically generates beacon frames after every Beacon Interval (BI) in order to identify its PAN to synchronize with associated nodes and to describe the superframe structure.
2) *Non-beacon-enabled mode:* all nodes can send their data by using an unslotted CSMA/CA mechanism, which does not provide any time guarantees to deliver data frames.

Superframe structure is only used in the beacon-enabled mode. The PAN coordinator uses it to synchronize associated nodes. A superframe is always bounded by two consecutive beacons and may consist of an active period and an optional inactive period, as shown in Figure 1. All communications must take place during the active part. In the inactive part, devices can be powered down/off to conserve energy.

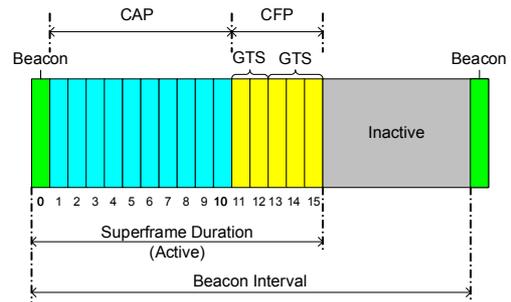

Figure 1. **Superframe structure**

The active part of the superframe is divided into 16 equally-sized slots and consists of 3 parts: a beacon, a contention access period (CAP) and an optional contention free period (CFP). The beacon shall be transmitted at the start of slot 0 without the use of CSMA/CA, and the CAP shall commence immediately after the beacon and complete before the beginning of CFP on a superframe slot boundary. In the CAP, slotted CSMA/CA is used as channel access mechanism. The CFP, if present, follows immediately after the CAP and extends to the end of the active portion of the superframe. In the CFP, CSMA/CA mechanism isn't used. Time slots are assigned by the coordinator for special applications such as low-latency applications or applications requiring specific data bandwidth. Devices which have been assigned specific time slots can transmit packets in this period. The specific time slots are called guaranteed time slots (GTSs). GTS can be activated by the request sent from a node to the PAN coordinator. Upon the reception of this request, the PAN coordinator checks whether there are sufficient resources available for the requested node to allocate requested time slot. A maximum of 7 GTSs can be allocated in one superframe. A GTS may occupy more than one slot period. Each device transmitting in a GTS shall ensure that its transaction is complete before the time of the next GTS or the end of the CFP. The allocation of the GTS can't reduce the length of the CAP to less than 440 symbols (aMinCAPLength).

The superframe structure is described by two parameters: beacon order (BO) and superframe order (SO). Both parameters can be positive integers between 0 and 14. The values of BO and SO are used to calculate the length of the superframe (i.e. beacon interval, BI) and its active period (i.e. superframe duration, SD) respectively, as defined in the following:

$BI = aBaseSuperframeDuration \times 2^{BO}$.
$SD = aBaseSuperframeDuration \times 2^{SO}$.
$Duty\ Cycle = SD / BI = 2^{SO-BO}$

where *aBaseSuperframeDuration*, a constant, describes the number of symbols forming a superframe when SO is equal to 0. The BO and SO must satisfy the relationship $0 \leq SO \leq BO = 14$. According to the IEEE 802.15.4 standard, the superframe will not be active anymore if *SO = 15*. Moreover, if *BO = 15* the superframe shall not exist and the non-beacon-enabled mode will be used. We use Duty Cycle show the relationship between BI and SD.

### 4.2 CSMA/CA Mechanism

In IEEE 802.15.4 standard, the channel access mechanism is often divided into slotted CSMA/CA for the beaconed-enabled mode and unslotted CSMA/CA for the non-beaconed-enabled mode, depending on network configurations. In both cases, the CSMA/CA algorithm is implemented based on backoff periods, where one backoff period shall be equal to a constant, i.e. aUnitBackoffPeriod (20 symbols). If slotted CSMA/CA is used, transmissions will be synchronized with the beacon, and hence the backoff starts at the beginning of the next backoff period. The first backoff period of each superframe starts with the transmission of the beacon, and the backoff will resume at the start of the next superframe if it has not been completed at the end of the CAP. In contrast, in the case of unslotted CSMA/CA, the backoff starts immediately. In the CSMA/CA algorithm each device in the network has three variables: NB, CW and BE.

- *NB* stands for the number of backoffs. It is initialized to 0 before every new transmission. Its maximum value is 4.
- *CW* means contention window and just exists in slotted CSMA/CA. It defines the number of backoff periods that need to be clear of channel activity before the transmission can start. It is initialized to 2 before each transmission attempt and reset to 2 each time the channel is accessed to be busy.
- *BE* is the backoff exponent. The backoff time is chosen randomly from [0, $2^{BE}$-1] units of time. The default minimum value (MinBE) is 3. The maximum value (MaxBE) is just 5, which prevents backoff delay time from becoming too long to affect the overall performance.

Each time a device needs to transmit data frames or MAC commands, it shall compute a backoff delay based on a random number of backoff period and performs CCA (clear channel assessment) before accessing to the channel. If the channel is busy, both NB and BE are incremented by 1, and CW is reset to 2. The device needs to wait for another random period and repeat the whole process. If the channel is sensed to be idle, CW is decreased by 1. And then if CW is equal to 0, the device can start transmit its data on the boundary of next backoff period. Otherwise the device needs to wait for another random period and repeat from CCA.

## 5. Simulation Settings

In this section we describe the configuration and settings of our simulation model in OMNeT++, including simulation scenario and parameter settings, and definition of performance metrics.

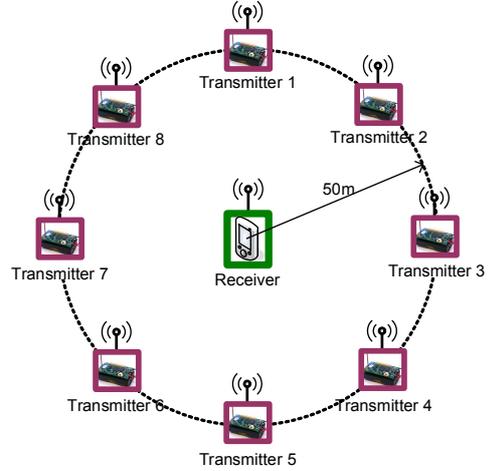

Figure 2. **Simulated network topology**

As mentioned previously, compared to peer-to-peer networks, star networks could be preferable for CPS applications and yield smaller delays because the communication in star networks occurs only between devices and a single central controller while any device in the peer-to-peer networks can arbitrarily communicate with each other as long as they within a common communication range. In this paper we focus on a one-hop star network, as shown in Figure 2. It consists of a number of transmitters and a central receiver. The transmitters are uniformly distributed around a 50-meter radius circle while the receiver is placed at the centre of the circle. The transmission range of every node is 176 m. Therefore we can easily learn that all the nodes are set to be in each other's radio range. Hence, there are no hidden nodes. The transmitters can be taken as devices such as sensors communicating to the central coordinator. The number of transmitters will change with scenarios in non-beacon mode. All transmitters periodically generate a packet addressed to the receiver. In the PHY layer, we use the 2.4 GHz range with a bandwidth of 250 kbps.

We select some important parameters, which may have significant influence on the performance of IEEE 802.15.4, as variable parameters, including MSDU (MAC service data unit) size, packet generation interval, MaxNB, MinBE and MaxFrameRetries in non-beacon mode; and MaxNB, BO and SO in beacon-enabled mode. They will be introduced with scenarios in the next two sections. Some important fixed parameters and default values of variable parameters are listed in Table 2.

As mentioned in Section 3, the performance of network protocols for CPS needs to be real-time, reliable, and resource efficient. In order to meet these requirements, we select end-to-end delay, effective data rate and packet loss rate as QoS metrics:

- *End-to-end delay:* it is a crucial metric to evaluate the real-time performance of networks. It refers to

the average time difference between the points when a packet is generated at the network device (transmitter) and when the packet is received by the network coordinator (receiver).
- *Effective data rate:* it is an important metric to evaluate the link bandwidth utilization which reflects the resource efficiency as well as dependability of networks. It is defined as below:

$$R_{effData} = N_{susspacket} \times L_{MSDU} / T_{end} - T_{start}$$

where $N_{susspacket}$ is the total number of usable date packets which are received successfully by coordinator from all devices in the simulation time. $L_{MSDU}$ is the MSDU length of the data frame. $T_{end} - T_{start}$ is the total time of the transmission from the beginning to the end.
- *Packet loss rate:* it indicates the performance of reliability, thus being an important metric. It is the ratio of the number of packets dropped by the network to the total number of packets generated at all devices.

From the above definitions, we can find that the effective data rate is closely related with the packet loss rate. Higher packet loss rate leads to lower effective data rate for the same number of transmitters. Hence in next section we sometimes analyze them together.

Table 2. **Parameter settings**

| Parameter | Value |
|---|---|
| Carrier Frequency | 2.4 GHz |
| Transmitter power | 1 mW |
| Carrier sense sensitivity | -85 dBm |
| Transmission range | 176 m |
| Bit rate | 250 Kbps |
| Traffic type | exponential |
| Number of packets sent by every device (in non-beacon enabled mode) | 5000 |
| Run time (in beacon-enabled mode) | 1000s |
| MaxBE | 5 |
| MinBE | 3 (default) |
| MaxNB | 4 (default) |
| MaxFrameRetries | 3 (default) |
| MAC payload size (MSDU size) | 60 Bytes (default) |
| Packet generation interval (in non-beacon enabled mode) | 0.025s (default) |
| Packet generation interval (in beacon- enabled mode) | 0.05s |
| Superfame order (SO) (in beacon- enabled mode) | 6 (default) |
| Beacon order (BO) (in beacon- enabled mode) | 7 (default) |
| Number of devices (in beacon- enabled mode) | 8 |

## 6. IEEE 802.15.4 in Non-beacon-enabled Mode

In the previous section, we have described the common settings for our simulations. This section will analyze the impact of five impact factors (i.e. MSDU size, packet generation interval, MaxNB, MinBE, and MaxFrameRetries) on the performance of IEEE 802.15.4 networks in terms of the above mentioned metrics respectively. During the process of simulation, when a specific parameter is examined as the impact factor, other parameters take the default values.

### 6.1 Impact of MSDU Size

MSDU size is the payload size of MAC layer and its maximum is 128 bytes. Fig. 3 shows its influence on the performance metrics for different number of transmitters.

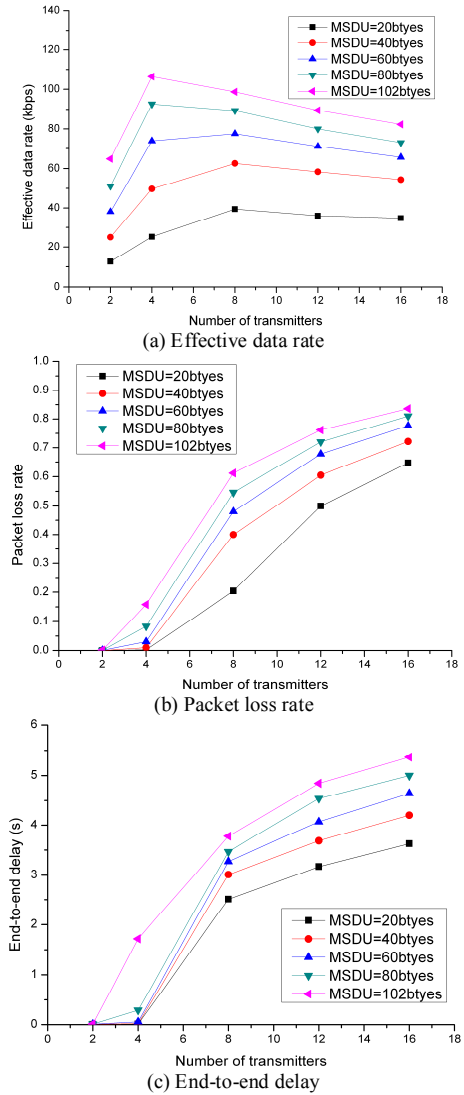

(a) Effective data rate

(b) Packet loss rate

(c) End-to-end delay

Figure 3. **QoS with different MSDU sizes**

Fig. 3(a) depicts the measured effective data rate, which increases with MSDU size for the same number of transmitters. This is because the effect of overhead was reduced, leading to a raise of data efficiency. We can also find that for a given MSDU size, when the number of transmitters increases, the effective data rate first increases and then decreases. This effect can be explained as follows. As the number of transmitters increases, more packets are sent in the same time, which cause the first increase of effective data rate. But too many packets will lead to packet collision and some conflicting packets are dropped. This is why the effective data rate decreases later.

Fig. 3(b) shows the measured packet loss rate. For the same MSDU size, the packet loss rate in denser network is higher. One reason may be that in denser sensor networks, more transmitters compete to access the channel. Consequently, the probability of packet collision becomes higher. For a certain number of transmitters, we can observe that larger MSDU sizes lead to higher packet loss rates.

Fig. 3(c) shows the measured end-to-end delay. The curve trend in the figure is similar with that in Fig. 3(b). From the above analysis of packet loss rate, we know that more transmitters and larger MSDU sizes increase the probability of packet collision. This can increase times of backoff and retransmission which are a considerable factor for longer delay. Therefore, the delay grows as the increase of the number of transmitters and MSDU size as shown in Fig. 3(c).

### 6.2 Impact of Packet Generation Interval

All transmitters periodically generate a packet addressed to the receiver. The time interval between two packets' generation is referred to as packet generation interval. It is apparent the packet generation interval is inversely proportional to traffic load. The result is shown in Fig. 4.

Fig. 4(a) shows the measured effective data rate. When the packet generation interval is less than 0.1s, as the number of transmitters increases, the effective data rate first grows and then decreases. The reason for this phenomenon is that as the number of transmitters increases, more packets are sent in the same time and traffic load increases; but overly-heavy traffic load leads to higher possibility of collision which causes the decrease of the effective date rate. On the other hand, when the interval is larger than 0.1s, although the number of transmitters increases, the traffic load is still very low. This is the reason why the effective data rate always keeps increasing as the number of transmitters increases.

Fig. 4(b) shows the measured packet loss rate, which is lower when the packet generation interval is larger than 0.1s. This is because larger packet generation intervals imply lighter traffic load and hence few collisions happen. On the other hand, when the packet generation interval is less than 0.1s, we can find that for a given small packet generation interval, the packet loss rate increases with the number of transmitters. In the meantime, for a certain number of transmitters, the packet loss rate increases as the interval decreases. This could be explained that smaller packet generation intervals mean heavier traffic load which increases the probability of packet collision.

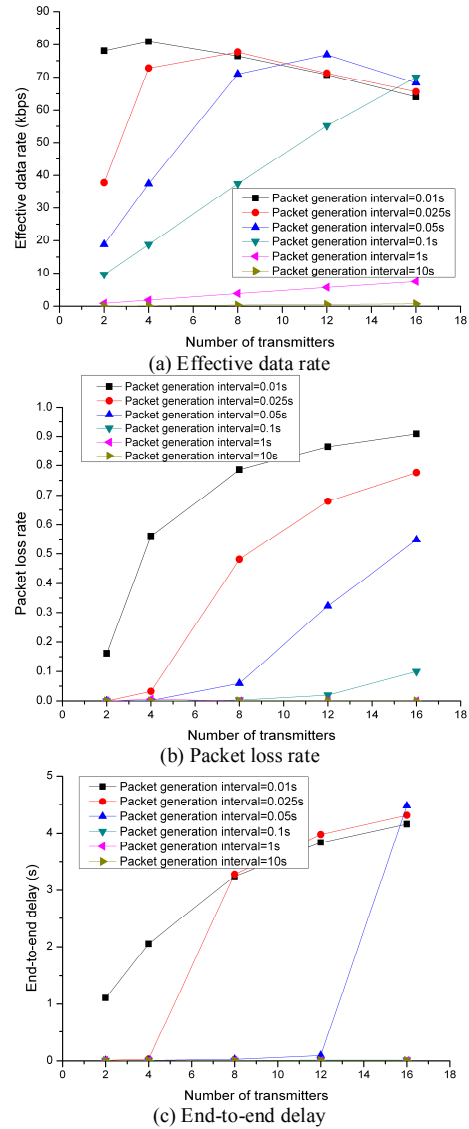

(a) Effective data rate

(b) Packet loss rate

(c) End-to-end delay

Figure 4. **QoS with different packet generation intervals**

Fig. 4(c) shows the measured end to end delay. We can see that when the packet generation interval is less than 1s, the end-to-end delay grows significantly with increasing number of transmitters. The reason for this is that for smaller packet generation intervals, the traffic load grows significantly as the number of transmitters increases. As a result, the competition of channel access is fierce and more backoffs and retransmissions are needed. On the other hand, when the packet generation interval is 1s or 10s, the end-to-end delay is close to zero and changes hardly as the number of transmitters increases.

## 6.3 Impact of MaxNB

MaxNB, as the name suggests, is the maximum number of CSMA backoffs. Its default value is 4. We vary it from 0 to 5. The result is given in Fig. 5. We can find that the default value of MaxNB is not the best selection.

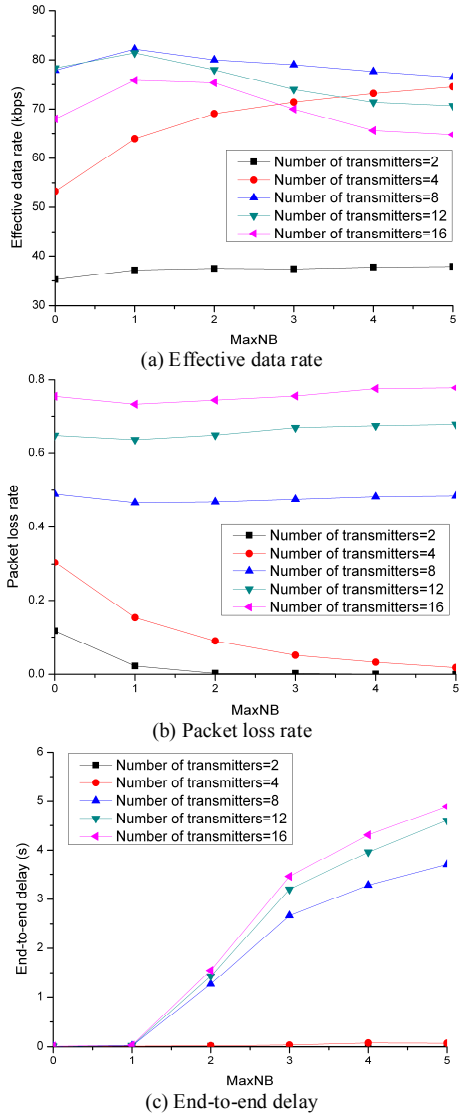

Figure 5. **QoS with different MaxNB values**

Fig. 5(a) shows the measured effective data rate, which grows for less (e.g. 4) transmitters as the value of MaxNB increase. However, when the number of transmitters reaches a certain threshold, the situation becomes opposite, as shown in the figure. In Fig. 5(b), for the same number of transmitters, contrary to the effective date rate in Fig. 5(a), the packet loss rate decreases for less transmitters with the increase of MaxNB. But when the number of transmitters reaches a certain threshold, the situation becomes opposite.

Fig. 5(c) shows the measured end-to-end delay, which is close to 0 for less (e.g. 2 or 4) transmitters as shown in the figure. This is due to that for less transmitters the channel is often idle and few collisions happen. On the other hand, for more transmitters, the delay grows with increasing MaxNB. This is because with increased number of transmitters, more times of backoffs will appear, which then lead to longer end-to-end delay.

## 6.4 Impact of MinBE

MinBE is the initial value of BE at the first backoff. Its default value is 3. We vary it from 1 to 5. The result is shown in Fig. 6.

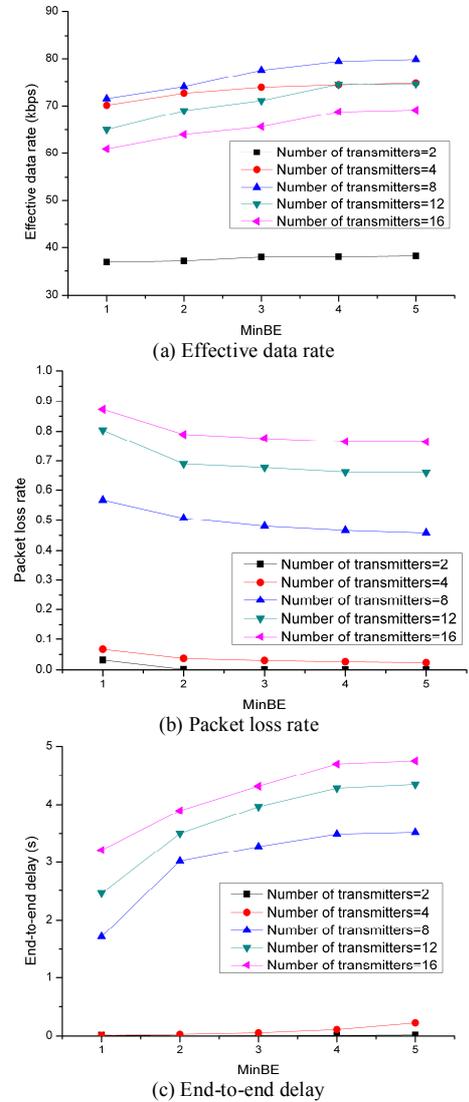

Figure 6. **QoS with different MinBE values**

Fig. 6(a) shows the measured effective data rate. We can observe that for the same number of transmitters, the effective data rate grows slowly as MinBE increases. Fig. 6(b) shows the measured packet loss rate, which decreases with the increase of MinBE and the number of transmitters. The reason for this may be that larger MinBE values imply larger backoff time, which cause the possibility of detecting an idle channel to increase. As a result, with the increase of MinBE, the effective data rate increases and the packet loss rate decrease for the same number of transmitters. Fig. 6(c) shows the measured end-to-end delay. At the same number of transmitters, the end-to-end delay grows with the increase of MinBE.

### 6.5 Impact of MaxFrameRetries

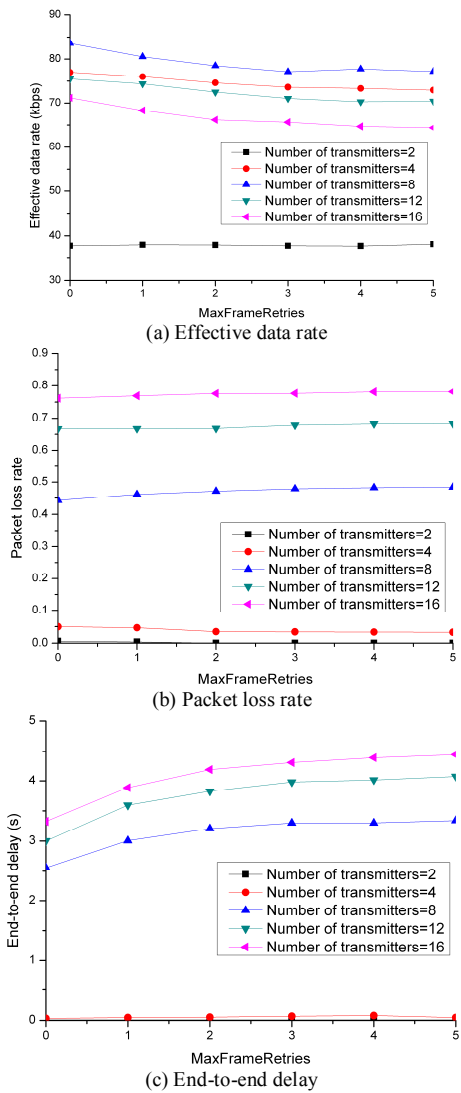

Figure 7. **QoS with different MaxFrameRetries values**

MaxFrameRetries refers to the maximum times of retransmission. If the retransmission times of a packet exceed the MaxFrameRetries value, it will be discarded. We vary MaxFrameRetries from 0 to 5. Fig. 7 shows the results.

Fig. 7(a) shows the measured effective data rate in this context. For a given larger number of transmitters, the effective data rate decreases slightly with the increase of MaxframeRetries while it increases for less transmitters. In Fig. 7(b), for the same number of transmitters the curve trend of packet loss rate is opposite to that of effective data rate in Fig. 7(a). The reason behind is similar with that of the MaxNB analysis for Fig. 5. Fig. 7(c) shows the measured end-to-end delay. We can learn that for less transmitters, the channel is often idle. Consequently, most of the frames can be transmitted successfully for the first time. As a result, the delay is close to 0. However, as the number of transmitters increases, the network load becomes heavier and the possibility of collision increases. Many packets need to be retransmitted for more times. This leads to the fact that end-to-end delay grows with the increase of MaxFrameRetries for the more transmitters.

To summarize the performance analysis in this section, in a network containing fewer transmitters, it is possible to improve its QoS by applying larger MSDU sizes and shorter packet generation intervals with tolerable delay. The MaxNB, MinBE, and MaxFrameRetries have less effect on sparse networks. On the other hand, in a dense network, with the same number of transmitters, the MSDU size and the packet generation interval are the main factors that influence the network QoS. Although MaxNB, MinBE, and MaxFrameRetries have less impact, it is possible to select appropriate values for them so that the performance of IEEE 802.15.4 can be improved, especially for reducing the mean end to end delay.

## 7. IEEE 802.15.4 in Beacon-enabled Mode

In this section, we analyze the performance of IEEE 802.15.4 in beacon-enabled mode. We will examine how MaxNB, SO and BO affect the network QoS with IEEE 802.15.4 standard in this context.

### 7.1 Impact of MaxNB

Here we examine the impact of MaxNB with different (BO, SO) values, with a duty cycle always equal to 50%. In this set of experiments, we vary MaxNB from 0 to 5.

Fig. 8(a) shows the measured effective data rate. Under the same duty cycle, it is clear that larger (BO, SO) values lead to larger effective data rates. This is because with smaller (BO, SO) values, beacons are transmitted more frequently. CCA deference is also more frequent in the case of lower SO values, which leads to more collisions at the start of each superframe. On the other hand, as the MaxNB value increases, the effective data rate increases gradually. This is due to larger MaxNB values that lead to higher probability of successful packet transmission.

Fig. 8(b) depicts the measured the packet loss rate. We observe that with the same BO value, a larger MaxNB can lead to a lower packet loss rate. On the other hand, with the same MaxNB, a smaller BO yields a higher packet loss rate. The reason for this phenomenon is that a larger MaxNB means a larger number of CSMA backoffs, resulting in more packets that can be transmitted successfully. In addition, a lower BO implies that beacons become more frequent. This is because the probability of packet collision becomes higher at the beginning of a new superframe.

Fig. 8(c) demonstrates the measured end-to-end delay. We can observe that with the same (BO, SO), the end-to-end delay increases with the value of MaxNB. This is because a larger MaxNB value implies a longer backoff time, which in turn may cause longer end-to-end delay. It can also be seen that for the same value of MaxNB, samller average delay can be obtained with larger (BO, SO) values. This is mainly due to the less packet collisions and retransmissions, which has been explained previously.

### 7.2 Impact of SO

As mentioned in Section 4, SO decides the length of superframe duration. In this subsection, we study its influence on network performance. The value of BO is set to 7. We vary SO from 1 to 6.

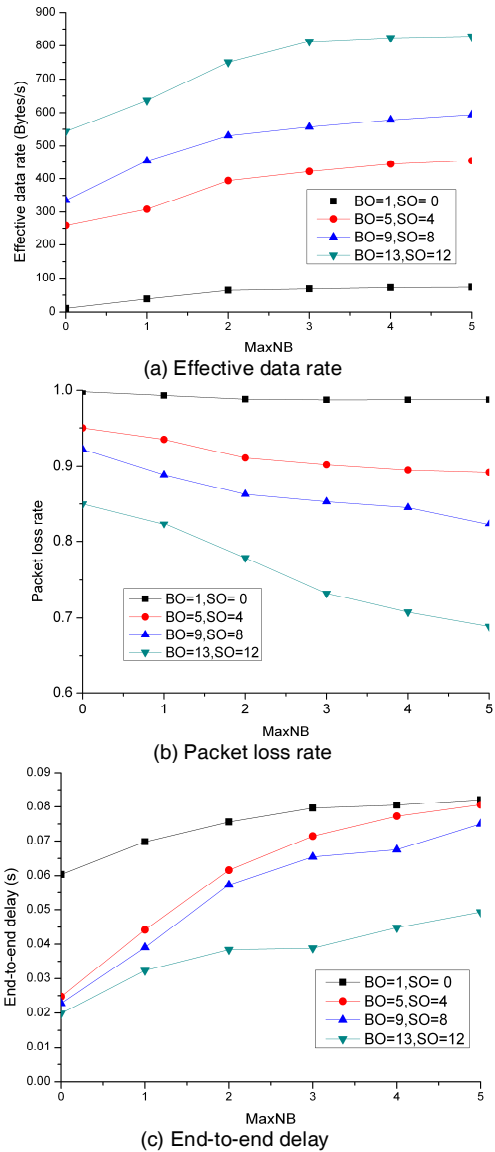

Figure 8. **QoS with different MaxNB values in beacon-enabled mode**

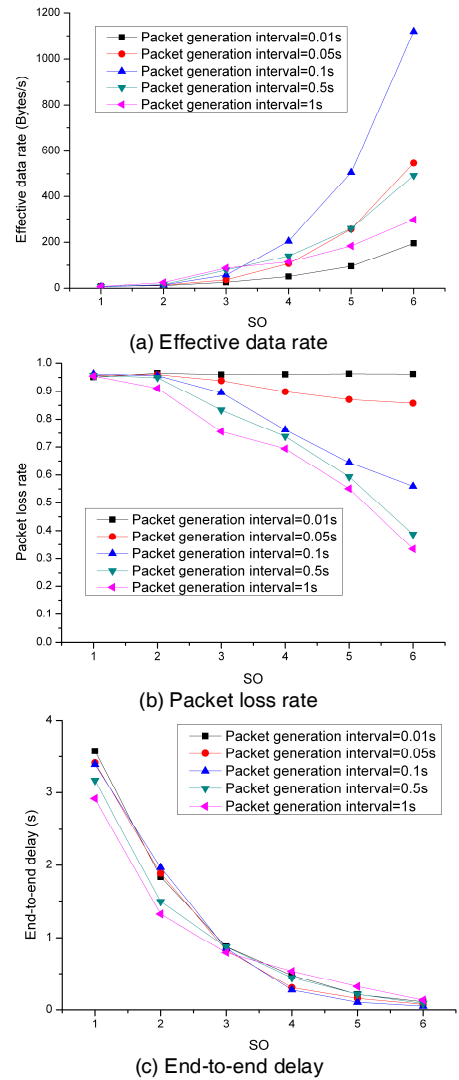

Figure 9. **QoS with different SO values**

Figs. 9(a) and 9(c) show the measured effective data rate and end-to-end delay, respectively. For the same packet generation interval, a larger SO with the same BO achieves a higher effective data rate and a lower end-to-end delay. This is because a larger SO implies a longer active period with a higher duty cycle. As a result, the network has better ability to transmit packets within current superframe, and hence less packets will experience a long sleeping delay.

Fig. 9(b) depicts the measured packet loss rate. It can be seen that with the same packet interval, a larger SO, which implies a higher duty cycle, yields a lower packet loss rate. When the packet interval is 0.01s, the packet loss rate is almost 100% all the time. The reason behind is that with a larger SO, more packets can be transmitted within the current superframe. On the other hand, with the same SO, the packet loss rate decreases as the packet generation interval increases.

### 7.3 Impact of BO

In this subsection, we examine the influence of BO on network performance. The value of BO controls the length of superframe (i.e. beacon interval). First, we fix the value of SO to 1. Then we examine network performance with different BO values. We vary BO from 7 to 2.

Fig. 10(a) shows the measured effective data rate. We can find that as the value of BO decreases, effective data rate grows gradually. This is mainly because the smaller BO resulting in higher duty cycle can achieve larger bandwidth, which implies larger effective data rates.

Fig. 10(b) gives the measured packet loss rate. It has been shown that for the same packet generation interval, a higher BO leads to a smaller packet loss rate. This is because under the same traffic load, the smaller BO resulting in larger duty cycle enables the network to transmit more packets. For the same BO, when the traffic load decreases, the packet loss rate descends from top (nearly 100%) to a very small value. This effect can be explained as follows: a smaller packet generation interval implies a higher traffic load and hence more packets need to be retransmitted as a result of collisions.

Fig. 10(c) presents the measured end-to-end delay. It is clear that higher delays are experienced for larger BO values with the same packet generation interval. The reason is that a larger BO causes a longer inactive period, in which case buffered packets may potentially experience a longer sleeping delay. For the same BO, the increase in packet generation interval results in decreased average delay. This is easy to understand since heavier traffic loads as a consequence of smaller packet generation intervals may cause more collisions and retransmissions.

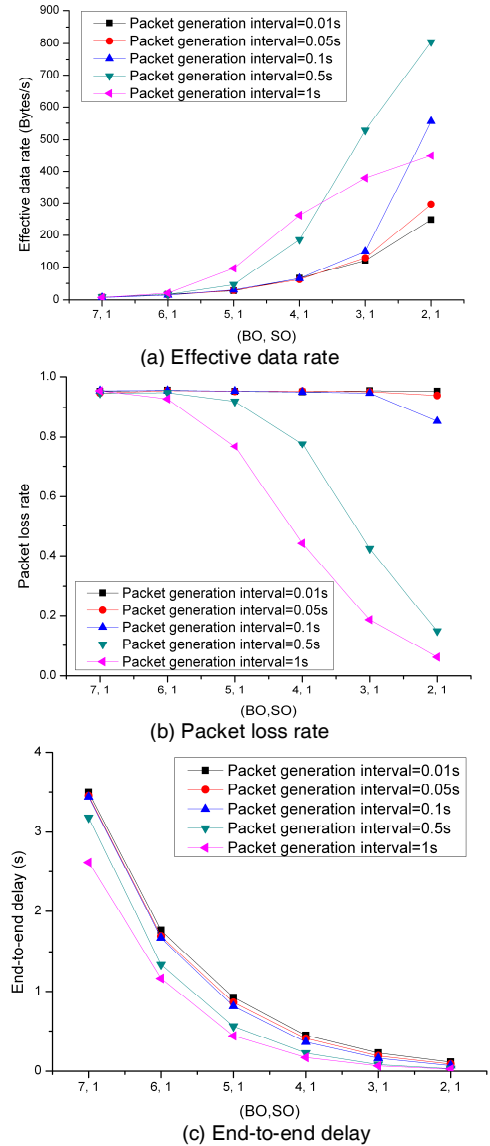

Figure 10. **QoS with different BO values**

To summarize this section, we extensively discussed a number of QoS measures of IEEE 802.15.4 standard in beacon-enabled mode. The results demonstrate the impact of MaxNB, BO and SO on the performance of the standard. By analyzing the results we can learn that in order to get low latency and high effective data rate, we should increase the value of SO and decrease the vlaue of BO as much as possible. We should select a suitable value for MaxNB, according to the requirements of the target CPS applications. In addtion, the results indicate that we should configure and optimize the protocol parameters by taking into account the practical application environments when designing CPS.

# 8. Conclusions

In this paper, we have presented a comprehensive performance evaluation of IEEE 802.15.4 standard in two different modes in the context of CPS. Considering general requirements of CPS applications, several network QoS metrics including effective data rate, packet loss rate, and end-to-end delay have been examined. We analyze them with respect to some important and variable protocol parameters. The analysis of simulation results provides some insights for configuring and optimizing the IEEE 802.15.4 protocol for CPS applications. A key finding is that the default configuration specified in the standard may not yield the best QoS in all cases. Consequently, some protocol parameters should adapt to the environments, while taking into account the CPS application requirements.

In future work, we will examine how to extend/modify IEEE 802.15.4 to make it more suitable for CPS applications. Self-adaptive and autonomous approaches will be our focus.

# Acknowledgment

This work was partially supported by the National Natural Science Foundation of China under Grant No. 60903153, the Fundamental Research Funds for the Central Universities (DUT10ZD110), Russian Foundation for Basic Research (project № 10-08-01071-a) and the SRF for ROCS, SEM, China.